\documentclass[showpacs,twocolumn,preprintnumbers,amsmath,amssymb,pra]{revtex4-2}
\usepackage{graphicx}
\usepackage{amsmath}
\usepackage{natbib}
\usepackage[dvipsnames]{xcolor}
\usepackage[colorlinks=true, allcolors=blue]{hyperref}
\usepackage{braket}
\usepackage{xcolor} 
\usepackage{placeins} %
\usepackage{float}

\begin{document}

\title{Multimode squeezed light generation and characterization}

\author{Alexander Chudakov$^{1,2}$}
\email{chudakov.as@phystech.edu}

\author{Vladislav Severin$^1$}

\author{Anastasia Poshevkina$^{1,2}$}

\author{Danil Malyshev$^{1,2}$}

\author{Emil Mingazhitdinov$^{3}$}

\author{Kirill Kuznetsov$^{1,3}$}

\author{Olga Tikhonova$^{1,3}$}

\author{Dmitry Kalashnikov$^{1,3}$}
\email{d.kalashnikov@rqc.ru}

\author{Polina R. Sharapova$^{1,3}$}
\email{p.sharapova@rqc.ru}

\affiliation{
$^{1}$Russian Quantum Center, Bolshoy Bulvar 30, bld. 1, 121205 Moscow, Russia
\\
$^{2}$Moscow Institute of Physics and Technology, 141700 Dolgoprudny, Russia
\\
$^{3}$Department of Physics, M.V.Lomonosov Moscow State University, Leninskie Gory, 119991 Moscow, Russia}

\begin{abstract}
Nowadays, the realization of quantum computations and communications based on continuous variables has attracted significant attention due to a substantial expansion of the  system dimensionality. The main progress in this area is attributed to the implementation of multimode systems based on squeezed states of light.  One of the simplest ways to generate such states relies upon their producing in a single-pass optical parametric amplifier (OPA) under ultrafast pumping. However, for homodyne detection of such multimode states, the profile of the local oscillator (LO) must perfectly match the profile of the measured mode. Usually, this is not the case; therefore, a proper treatment of multimode squeezing is required.
In this work, we study both  theoretically and experimentally  the multimode squeezed light generated in type-0 and type-II OPA. 
We characterize such sources using the Schmidt-mode representation and investigate the degree of squeezing depending on the LO spectral profile, employing a pulse shaping technique. In addition, we investigate cluster states that can be generated under the given experimental conditions.  
This work might have a significant impact on the realization of multimode quantum  protocols.

\end{abstract}

\maketitle

\section{Introduction}
\label{Introduction}
Since the first demonstration in mid-80s \cite{PhysRevLett.56.788} the squeezed states of light have become a vital resource for a number of quantum applications including quantum metrology and sensing, quantum communications and computation \cite{RevModPhys.77.513, RevModPhys.84.621, Andersen_2016, doi:10.1021/acsphotonics.9b00250, dcfq-1byk}. Indeed, reducing the quantum noise below the classical (shot-noise) limit in one of the quadratures opens up the rich plethora of quantum phenomena inaccessible with classical light. Although the generation of squeezed states with a continuous-wave (CW) pump is studied extensively, in recent years the alternative schemes with pulsed pump have attracted significant attention as they benefit in frequency and temporal multimode operations \cite{RevModPhys.92.035005, malik2026, he2026experimentaldetectionentanglementmultimode}. 
Such schemes allow to realize large multidimensional entangled states, which are the essential component for the continuous-variable (CV) based quantum  technologies such as measurement-based quantum computations (MBQC), multiparty communications and quantum-enhanced machine learning protocols \cite{PhysRevLett.112.120505, NatCom.8.15645, NatPhoton.7.982, doi:10.1126/science.aay2645, doi:10.1126/science.aay4354, PhysRevA.100.022303, Nokkala_2018, CommunPhys.4.53, yokoyama2026experimentalimplementationcontinuousvariableqaoa}.

The experimental realization of such multidimensional entangled states requires two important and mutually linked steps, i.e. efficient generation of multimode squeezed light and its subsequent detection. Among the proposed schemes for generating multimode squeezed light, the single-pass optical parametric amplifier (OPA) configuration appears to be the simplest one, as it does not require complex cavity alignment  \cite{RevModPhys.92.035005,he2026experimentaldetectionentanglementmultimode, NatPhys.16.144, Kouadou23, PhysRevResearch.6.043113, NatPhoton.19.526, NatPhoton.20.156, Kalash2026}. In this  scenario, the generated squeezed light can be represented in the basis of broadband Schmidt modes \cite{RevModPhys.92.035005, NatPhoton.8.109},  the number and structure of which depend on the crystal and pump parameters that ensure
control of the mode structure through dispersion engineering and pump shaping
\cite{PhysRevA.73.063819, PhysRevA.74.061801, NatPhoton.8.109, Ansari:18, PhysRevA.97.033808}. The detection of squeezed light constitutes another important aspect and usually involves the balanced homodyne detection. In this scheme,  the strong coherent light (local oscillator, LO) is superimposed with the squeezed light in a beamsplitter
with a subsequent measurement of the photocurrent difference by a balanced photodetector \cite{andrews2015photonics}. 
However, in the case of multimode squeezed light, homodyne detection proves to be less straightforward as in a single-mode case, since LO must match spectral, spatial and phase properties of the light under study \cite{PhysRevA.73.063819}. Moreover, MBQC schemes relying upon the Schmidt-mode decomposition of squeezed light  require precise matching between the LO and Schmidt modes for correct processing 
\cite{Kouadou23, PhysRevResearch.6.043113, NatPhoton.8.109}. However, preparing LO with a specific Schmidt-mode profile often presents a non-trivial task. In this regard, the question arises as to how the measurement outcomes depend on the LO profile in cases of imperfect matching.

In this work, we study both theoretically and experimentally the influence of the LO frequency profile on the measured squeezing  of multimode squeezed light produced in the single-pass OPA scheme. We consider both type-0 and type-II parametric down-conversion  (PDC) and present a detailed characterization of the multimode squeezed light sources, which includes numerical modeling of the joint spectral amplitudes and corresponding Schmidt modes, calculating and measuring spectral intensity distributions together with squeezing in various modes. We also analyzed a possible  generation of cluster states under the given experimental conditions.
Such a  comprehensive analysis  constitutes a key aspect for the realization of multidimensional entangled states and protocols based on the multimode squeezed light.

\section{Theoretical model}
\label{Theory}

\subsection{Multimode Parametric Down-Conversion}

In this section, we briefly remind the Schmidt-mode theory \cite{PhysRevLett.84.5304,  PhysRevA.97.053827} used for describing multimode PDC process \cite{hong1985theory, klyshko2018photons}. We start our theoretical description by considering a frequency-degenerate collinear PDC  occurred in a nonlinear periodically poled crystal \cite{161322, CRPHYS_2007__8_2_180_0} of length $L$, 
which effective Hamiltonian in the interaction picture under the undepleted pump approximation can be written as 

\begin{equation}
\hat{H} = i\hbar \tilde{\Gamma} \int d\omega_s\, d\omega_i\, F(\omega_s, \omega_i)\, \hat{a}^\dagger (\omega_s) \hat{b}^\dagger(\omega_i) + \mathrm{H.c.},
\label{eq: Hamiltonian1}
\end{equation}
where $\hat{a}^\dagger(\omega_s), \hat{b}^\dagger(\omega_i)$ are the creation operators for the signal (s) and idler (i) photons at frequencies $\omega_s $ and $ \omega_i$, respectively; $\tilde{\Gamma}$ is the coupling strength that quantifies the parametric gain \cite{PhysRevA.73.063819, klyshko2018photons} and depends on the parameters of the pump and nonlinear medium. The function $F(\omega_s,\omega_i)$ is the normalized joint spectral amplitude (JSA) \cite{PhysRevA.56.1627}, which for a pulsed pump field with a Gaussian temporal envelope and central frequency $\omega_p$  can be written as

\begin{equation}
\begin{aligned}
F(\omega_s, \omega_i) = &\frac{1}{N} \exp\left(-\frac{(\omega_p - \omega_s - \omega_i)^2 \tau^2}{2}\right) \\
&\times \mathrm{sinc}\left(\frac{\Delta\beta L}{2}\right) \exp\left[i \frac{\Delta\beta L}{2}\right],
\end{aligned}
\label{eq: JSA1}
\end{equation}
where $N$ is the normalization constant, $\tau$ is the characteristic pump parameter related to the pump pulse duration as  $\tau_p=2\sqrt{ln2}\tau$. The phase-mismatch is defined as $\Delta \vec{\beta} = \vec{k}_p - \vec{k}_s - \vec{k}_i - \frac{2\pi}{\Lambda}=0$, where  $\vec{k}_{i,s,p}$  denote the  wave vectors of the pump, signal, and idler fields, respectively, and $\Lambda$ is the poling period of the crystal.

The JSA can be decomposed using the Schmidt decomposition \cite{PhysRevLett.84.5304} 
\begin{equation}
F(\omega_s, \omega_i) = \sum_n \sqrt{\lambda_n} \, u_n(\omega_s)\, v_n(\omega_i),
\label{eq: JSA schmidt decomposition}
\end{equation}
where $\lambda_n$ are the Schmidt eigenvalues, while  $u_n(\omega_s)$ and $v_n(\omega_i)$ constitute the orthonormal sets of functions (Schmidt modes) describing the spectral properties of the signal and idler fields, respectively. Such a decomposition allows us to introduce the Schmidt operators
\begin{equation}
\hat{A}_n^\dagger = \int d\omega_s\, u_n(\omega_s)\, \hat{a}^\dagger(\omega_s), \ \
\hat{B}_n^\dagger = \int d\omega_i\, v_n(\omega_i)\, \hat{b}^\dagger(\omega_i)
\label{eq: Operators}
\end{equation}
that diagonalize the PDC hamiltonian 
\begin{equation}
\hat{H} = i \hbar \Gamma  \sum_n  \sqrt{\lambda_n} \left(\hat{A}_n^\dagger \hat{B}_n^\dagger - \hat{A}_n \hat{B}_n\right).
\label{eq: Hamiltonian2}
\end{equation}
The input-output relations for these operators read
\begin{equation}
\begin{aligned}
\hat{A}_n^{out} = \hat{A}_n^{in} \mathrm{cosh}[\sqrt{\lambda_n} \Gamma ] + [\hat{B}_n^{in}]^{\dagger} \mathrm{sinh}[\sqrt{\lambda_n} \Gamma ],
\\
[\hat{B}_n^{out}]^{\dagger}= [\hat{B}_n^{in}]^{\dagger} \mathrm{cosh}[\sqrt{\lambda_n} \Gamma ] + \hat{A}_n^{in} \mathrm{sinh}[\sqrt{\lambda_n} \Gamma ],
\end{aligned}
\label{eq: input_output}
\end{equation}
where $\Gamma = \tilde{\Gamma} \ T $, and $T$ is the interaction time. Note that each pair of Schmidt modes with the same index $n$ undergoes its own squeezing
with the squeezing parameter  
$r_n =\sqrt{\lambda_n} \Gamma$ \cite{PhysRevA.91.043816}.
The output intensity distribution of the signal (analogously for idler) photons is given by
\begin{equation}
N(\omega_s) =  \sum_n  |u_n(\omega_s)|^2 \mathrm{sinh}[\sqrt{\lambda_n} \Gamma ]^2.
\label{eq: intensity_distr}
\end{equation}

\subsection{Squeezing measurement with broadband local oscillator}

Squeezing is usually measured using the homodyne detection technique, in which the strong LO of amplitude $\alpha_{\mathrm{LO}}$ interferes with the squeezed light at a beam splitter followed by the measurement of the photocurrent difference by a balanced photodetector, which, in its turn, is proportional to the quadrature of  the light $X_{\phi}$ under study: $\delta I \sim  \alpha_{\mathrm{LO}} X_{\phi}$ \cite{RevModPhys.81.299}. 
In the case of multimode squeezed light, measuring the squeezing in the selected Schmidt mode requires precise adjustment of both the spectral and phase profiles of the LO to match this mode. Practically, this can be realized with the pulse-shaping technique using a spatial light modulator (SLM). If this matching is not achieved, the LO field constitutes a superposition of multiple Schmidt modes and requires a proper description, which is presented below.   

\subsubsection{Type-0 and type-I PDC}
In the case of both type-0 and type-I PDC, the signal and idler Schmidt modes of the same index $n$ have the same spectral profiles: $u_n(\omega)=v_n(\omega)$, therefore the LO field can be decomposed with respect to the Schmidt-mode basis as
\begin{equation}
\psi_{\mathrm{LO}}(\omega) = \sum_{n=0}^{\infty} M_n e^{i\theta_n} u_n(\omega),
\label{eq: Psi_LO}
\end{equation}
where $M_n$ and $\theta_n$ are the amplitude and phase of the complex expansion coefficients, respectively, satisfying $\sum_{n=0}^{\infty} M_n^2 = 1$. 

Homodyne detection projects the state of quantum light onto the LO mode $\psi_{LO}(\omega)$, resulting in the measurement of squeezing of the generalized quadrature 
\begin{equation}
\hat{X}_{LO} = \frac{e^{i \phi} \hat{A}_{LO} - e^{-i \phi} \hat{A}_{LO}^{\dagger}}{2i}
\end{equation}
associated with the LO mode
\begin{equation}
\hat{A}_{LO}^\dagger= \int d\omega\, \psi_{LO}(\omega)\, \hat{a}^\dagger(\omega),
\end{equation}
where $\phi$ is the frequency independent phase of the LO.
 Taking into account the input-output relations for the Schmidt operators Eq. (\ref{eq: input_output}) and assuming the following commutation relation $[\hat{A}_n \hat{B}^{\dagger}_m]= \delta_{nm}$ for type-0 (I) PDC, one can calculate the variance of the generalized quadrature $ \Delta X_{LO,}=\langle\hat{X}^2_{LO,} \rangle - \langle\hat{X}_{LO,} \rangle^2 $, which includes the contributions from all the Schmidt modes and is given by \cite{PhysRevA.73.063819} 
\begin{equation}
\Delta X_{LO} = \sum_{n=0}^\infty \frac{M_n^2}{4} \left( e^{2r_n} \sin^2 (\theta_n+\phi) + e^{-2r_n} \cos^2 (\theta_n+\phi) \right).
\label{eq: Q_var1}
\end{equation} 
Note that if the LO spectral profile coincides with the Schmidt mode profile of order $m$,  the quadrature squeezing of this mode $m$ will be measured. 

\subsubsection{Type-II PDC}

In the case of type-II PDC, the signal and idler photons have different polarizations, and, therefore,  have different spectral profiles: $u_n(\omega) \neq v_n(\omega)$. This means that, in general, measurement of squeezing for type-II PDC requires two LOs, $\psi_{LO} (\omega)$ and $\varphi_{LO}(\omega)$, where $\psi_{LO} (\omega)$ has the polarization of the signal photon and is decomposed with respect to its modes according to  Eq.  (\ref{eq: Psi_LO}), while $\varphi_{LO}(\omega)$ has the polarization of the idler photon and is decomposed with respect to the idler Schmidt modes as  
\begin{equation}
    \varphi_{\mathrm{LO}}(\omega) = \sum_{n=0}^{\infty} C_n e^{i\chi_n} v_n(\omega),
   \label{eq: varphi_LO}
\end{equation}
where $C_n$ and $\chi_n$ are the amplitude and phase of the complex expansion coefficients, respectively, satisfying $\sum_n  C_n^2 = 1$.

Then, the generalized bipartite quadrature  can be defined as 
\begin{equation}
\hat{X}_{LO}=\frac{\hat{X}_{LO,\psi}+\hat{X}_{LO,\varphi}}{\sqrt{2}},
\end{equation}
where
\begin{equation}
\hat{X}_{LO,\psi}= \frac{e^{i \phi}\hat{A}_{LO} - e^{-i \phi}\hat{A}_{LO}^{\dagger}}{2i}
\end{equation}
and
\begin{equation}
\hat{X}_{LO,\varphi}= \frac{e^{i \phi}\hat{B}_{LO} - e^{-i \phi}\hat{B}_{LO}^{\dagger}}{2i}
\end{equation}
are the generalized  quadratures associated with the LO modes 
\begin{equation}
\hat{A}_{LO}^\dagger= \int d\omega\, \psi_{LO}(\omega)\, \hat{a}^\dagger(\omega), \   \hat{B}_{LO}^\dagger= \int d\omega\, \varphi_{LO}(\omega)\, \hat{b}^\dagger(\omega)
\end{equation}
of the signal and idler subsystems, respectively. Note that in our description, both LO profiles include the same LO phase  $\phi$.

Finally, taking into account the input-output relations for the Schmidt operators Eq. (\ref{eq: input_output}) and assuming the commutation relation $[\hat{A}_n \hat{B}^{\dagger}_m]= 0$ for type-II PDC, the variance of the generalized quadrature can be expressed as
\begin{multline}
\Delta X_{LO} = \\
\frac{1}{16} \sum_{n=0}^\infty \big[ e^{2r_n} \big(M_n^2 + C_n^2 - 2 M_n C_n \cos{(\theta_n + \chi_n +2 \phi)}\big)
\\
+ e^{-2r_n} \big(M_n^2 + C_n^2 + 2 M_n C_n \cos{(\theta_n + \chi_n  +2 \phi)}\big) \big],
\label{eq:Q_var type-2}
\end{multline}
where  $\Theta_n=\theta_n + \chi_n$ represents a joint phase.
If the LO spectral profiles of the signal and idler subsystems coincide with the signal and idler Schmidt modes of order $m$, the bipartite squeezing of these modes will be measured.

\subsubsection{Losses}

Taking the same amount of losses $R$ for both the signal and idler subsystems, the quadrature variance can  be written as (see Appendix \ref{app: C})
\begin{equation}
\Delta X_{LO}^{(loss)} = (1 - R) \Delta X_{LO} + R \Delta X_{vac},
\end{equation}
whether $\Delta X_{LO}$ is given by Eq. \eqref{eq: Q_var1} or Eq. \eqref{eq:Q_var type-2} for PDC type-0 (I) and type-II, respectively, $\Delta X_{vac}=\frac{1}{4}$ is the vacuum noise level (the variance of the vacuum field).
Then the multimode squeezing parameter can be calculated as
\begin{equation}
    r_{\text{dB}} = 10 \log_{10}\left(\dfrac{\Delta X_{LO}^{(loss)}}{\Delta X_{vac}}\right).
    \label{squeezing}
\end{equation}

\section{Experimental setup}
\label{Experimental setup}

\begin{figure}[h]
  \centering
  \includegraphics[width=0.5\textwidth]{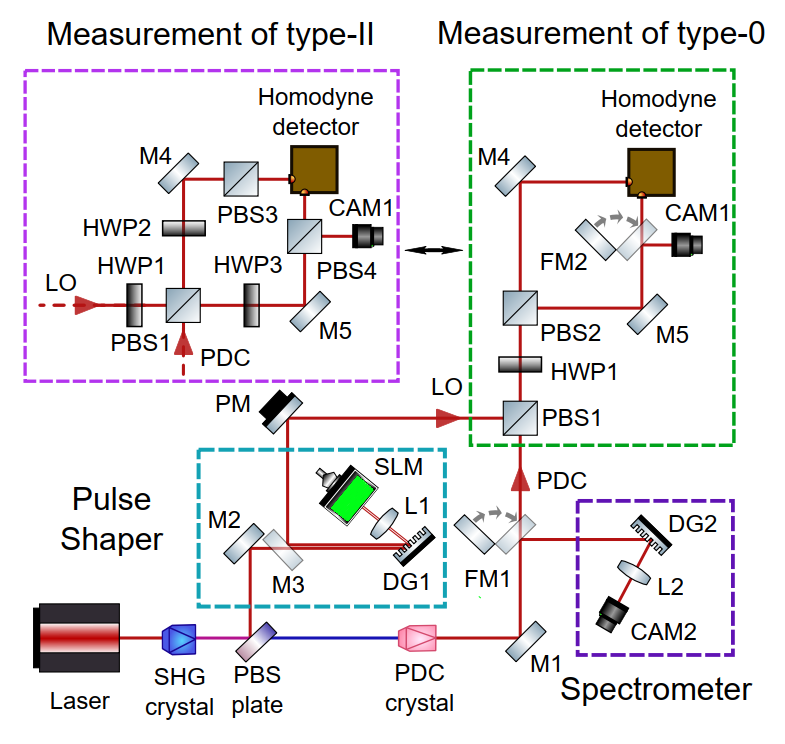}
  \caption{The sketch of the experimental setup includes three main blocks: Pulse Shaper, Spectrometer and Squeezing Measurement. The choice of the system configuration depends on the type of PDC process (0 or II). Here, Laser is the picosecond Ti: Sapphire Coherent Mira 900D pulsed laser, SHG crystal is the crystal for the second harmonic generation, PBS plate is the polarizing beam-splitting plate, PDC crystal is the crystal for producing the type-0(II) PDC process, M(1-5) are mirrors, DG1(2) are diffraction gratings, L1(2) are lenses, FM1(2) are flip mirrors, SLM is the spatial light modulator, PM is the piezo-mirror, PBS(1-4) are  polarizing beam-splitting cubes, HWP(1-3) are half-wave plates, CAM1(2) are CMOS cameras and Homodyne detector is a device for homodyne measurements.
  The system of PBS3(4) and HWP2(3) is necessary for bringing LO and squeezed light to the same polarization mode.}
  \label{fig: Setup}
\end{figure}

Fig. \ref{fig: Setup} shows a simplified optical scheme of the experimental setup. The setup is based on the picosecond Ti: Sapphire Coherent Mira 900D pulsed laser with an average output power of $0.9\ \text{W}$, the pulse repetition rate of approximately $76\ \text{MHz}$, the pulse duration of about $2\ \text{ps}$ and with the central wavelength of about $783.3\ \text{nm}$.

The radiation of the Ti:Sapphire laser produces the second harmonic generation (SHG, type-I) process at the wavelength of $391.65\ \text{nm}$ in the lithium triborate (LBO) crystal with the length of $17\ \text{mm}$. The polarizing beam-splitting plate (PBS-plate) separates the remaining laser radiation from the SHG (here and after referred to as the pump), the remaining laser radiation is later used as the local oscillator. The pump with an average power of about $40\ \text{mW}$ passes through the periodically poled potassium titanyl phosphate (PPKTP) crystal with the length of $1\ \text{mm}$ for type-0 (Raicol, polling period $2.95\ \mu m$) and $2\ \text{mm}$ for type-II (Raicol, polling period $7.95\ \mu m$) PDC (PDC crystal), resulting in the generation of the multimode squeezed light. To avoid the “gray track” effect, the PPKTP crystals were set to operate at the temperature above $100^\circ\text{C}$ ($T = 105^\circ\text{C}$ for type-0, and $T = 103^\circ\text{C}$ for type-II PDC, respectively) \cite{Boulanger, Motokoshi:01}. Reflecting from the mirror M1, the generated multimode squeezed light impinged on the polarizing beam splitter PBS1, where it was superimposed with the LO.

The PDC spectra for both type-0 and type-II PDC were characterized by the self-made spectrometer consisting of the diffraction grating DG2, the focusing lens L2, and the camera CAM2 (Thorlabs CS135).

The desired spectral and spatial profiles of the LO were set by the Pulse Shaper, which consists of the SLM (Holoeye LETO II, 1080x1920 pixels resolution) introduced into the folded 4f-system with DG1 diffraction grating (Thorlabs GR25-1208, optimized for a wavelength of 750 nm, 1200 grooves/mm) and the L1 lens (focal length of 500 mm). This scheme converts the frequency spectrum into its spatial counterpart, sets a phase mask to the beam in the Fourier plane using the SLM, and then transforms the resulting spatial spectrum back into the frequency domain. Here, we used the well-known procedure for generating phase masks described in \cite{Bolduc:13,Bobrov:15,LaVolpe:20}. More details about Pulse Shaping can be found in Appendix~\ref{app: D}. The relative phase $\phi$ between the LO and the squeezed light was swept using the piezo mirror PM introduced into the optical path of the LO, which was driven by the sawtooth signal (peak voltage $20\ \text{V}$, frequency $1\ \text{Hz}$).

In both cases, i.e. type-0 and type-II PDC, the polarization state of the multimode squeezed light differs from its LO counterpart.  We bring them to the same polarization by the set of polarization beamsplitters and half-wave plates before the home-made homodyne detector (bandwidth of about 80 MHz, shot-noise clearance (SNC) of 13 dB and the common-mode rejection ratio (CMRR) value of 52.4 dB \cite{Kumar_2012}), though at the expense of additional losses (for more information, see the measurement blocks in Fig. \ref{fig: Setup}). The measured  photocurrent difference was processed using the Analog-to-digital converter (Agilent U1084A) integrated into the personal computer. Subsequently, the variance of the collected photocurrent statistics was calculated by the developed software, allowing to estimate the squeezing value of the measured quadrature.

\section{Results}

\subsection{Spectra and Schmidt modes}

\begin{figure*}
    \centering
    \includegraphics[width=1\textwidth]{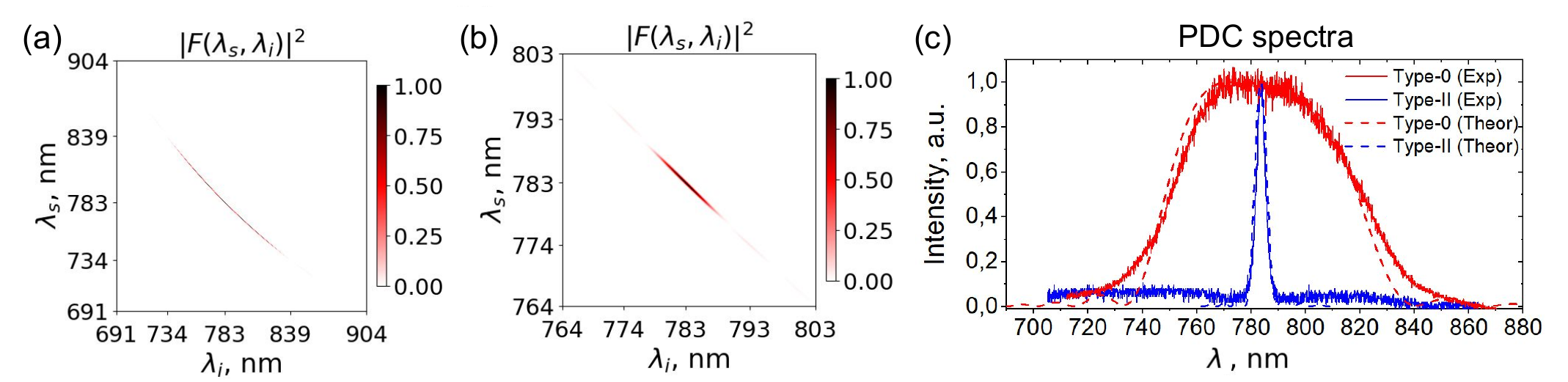}
    \caption{Joint spectral intensities normalized to unity at their maximum for the (a) type-0 PDC and (b) type-II PDC processes. (c) Normalized theoretical (dashed) and experimental (solid) PDC spectra for the type-0 (red) and type-II (blue) PDC processes.}
    \label{fig:JSI}
\end{figure*}

For numerical simulations, we used parameters of the experimental setup described in Sec. \ref{Experimental setup}. The substantial experimental heating of the nonlinear crystals was  accounted for using the appropriate dispersion relations \cite{Emanueli}.
Since the length of the nonlinear crystal used in the experiment was quite large, the effective interaction length was considered to be limited 
 by the pump coherence length ($L_{\text{coh}}$): $L_{\text{eff}} = \dfrac{L_{\text{coh}}}{n_p(\omega_p)}$, where $n_p$ is the pump refractive index.  For the type-0 PDC, the estimated effective length equals to $L_{\text{eff}} = 0{.}8077$ mm, whereas for the type-II PDC it constitutes $L_{\text{eff}} = 0.8634$ mm. 
The corresponding Joint Spectral Intensities (JSIs) are shown in Fig. \ref{fig:JSI} (a,b). Note that when transforming the JSI and intensity distribution from the frequency to the wavelength domain, both the spectrum and the JSI must be recalculated according to the corresponding  Jacobian $\big| \frac{\text{d}\omega}{\text{d}\lambda} \big| = \frac{2\pi c}{\lambda^2}$ to maintain the normalization.  By integrating the JSI over the idler (signal) frequency, we  obtain the signal (idler) intensity distributions 

\begin{equation}
I(\omega_s) = \displaystyle\int \left|F(\omega_i,\omega_s)\right|^2 d\omega_i, \ \
I(\omega_i)= \displaystyle\int \left|F(\omega_i,\omega_s)\right|^2 d\omega_s,
\label{eq: Theor spectra}
\end{equation}
which in our case (due to small gain) coincide with $N(\omega_s)$ and  $N(\omega_i)$, respectively, given by Eq. (\ref{eq: intensity_distr}).

\begin{figure*}
  \centering
  \includegraphics[width=0.95\textwidth]{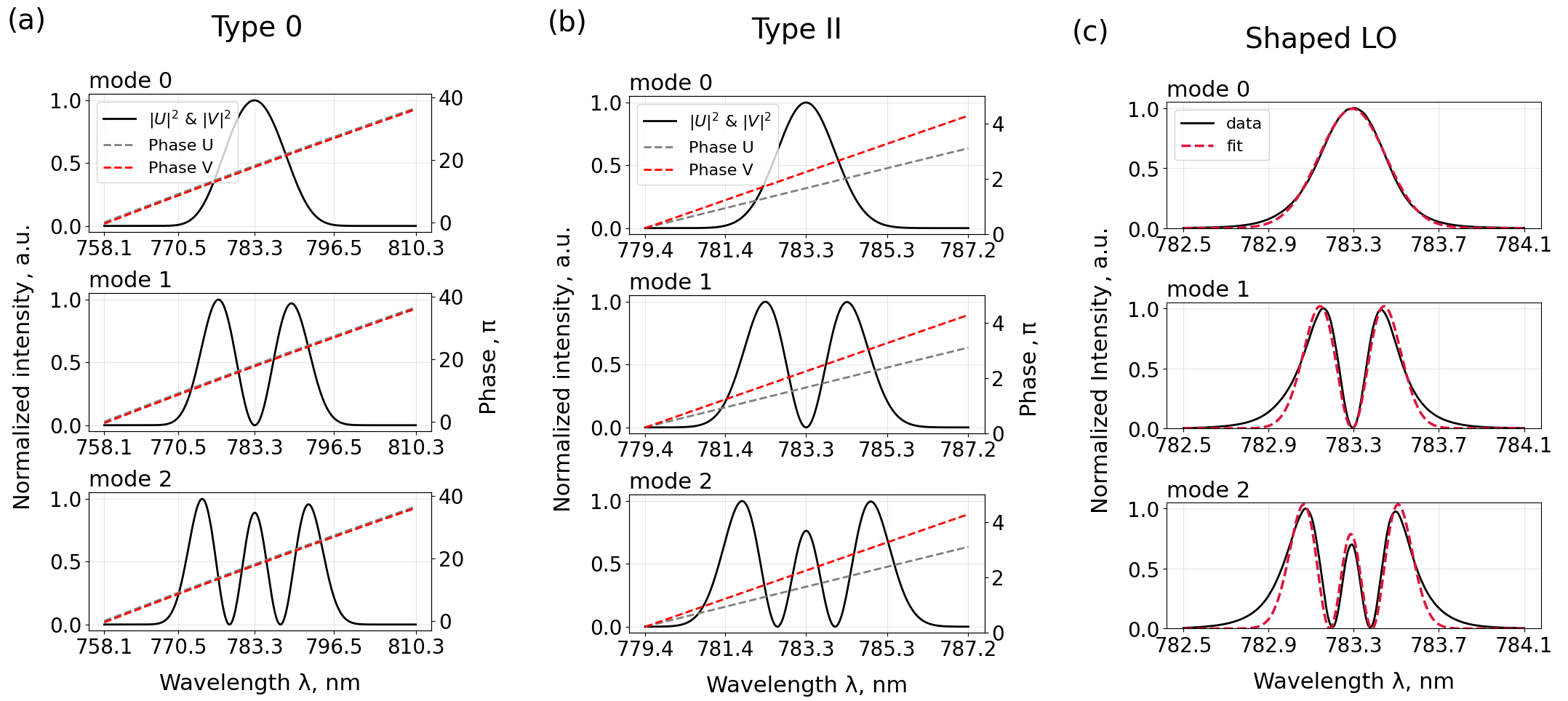}
  \caption{Schmidt modes normalized to unity at their maximum value: (a) for the type-0 PDC process and (b) for the type-II PDC process. (c) Spectra of the local oscillator after pulse shaping (black) and the corresponding fitting curves (red). The fitting was performed using the squared Hermite–Gaussian functions. 
  \label{fig: Schmidt modes}}
\end{figure*}

Fig. \ref{fig:JSI} (c) shows a comparison of the numerically calculated spectra according to Eq. (\ref{eq: Theor spectra}) and the experimentally measured spectra 
for type-0 and type-II PDC. Note that for our source, due to the JSA symmetry, the signal and idler spectra coincide.  Experimental spectra were measured using the self-made spectrometer calibrated with the tunable narrow-band laser. As it can be seen, the experimental and theoretical spectra show a good agreement for both type-0 and type-II PDC.

The Schmidt decomposition of the JSA  (see Eq. (\ref{eq: JSA schmidt decomposition})) has been performed numerically using the singular value decomposition (SVD). 
The amplitudes of the Schmidt modes and their phase distributions are shown in Fig. \ref{fig: Schmidt modes} (a,b), the corresponding eigenvalue distributions for both PDC types are given in Appendix \ref{appendix: Schmidt eigenvalues}. 
Note that in the case of periodically poled crystals, the Schmidt-mode phase depends linearly on the wavelength for both PDC types, since the first order dispersion term dominates in the wavevector decomposition. This leads to the separation of the signal and idler variables in the phase-mismatch  of Eq. \eqref{eq: JSA1} for both type-0 and type-II PDC:
\begin{equation}
\begin{aligned}
\Delta \beta \approx  &\big( \frac{\text{d}k_p}{\text{d}\omega}\big|_{\omega_p} - \frac{\text{d}k_s}{\text{d}\omega}\big|_{\omega_p/2} \big) \big( \frac{\omega_p}{2} - \omega_s\big)
\\[2mm]
&
+ \big( \frac{\text{d}k_p}{\text{d}\omega}\big|_{\omega_p} - \frac{\text{d}k_i}{\text{d}\omega}\big|_{\omega_p/2} \big) \big( \frac{\omega_p}{2} - \omega_i\big)
\\
&=\Delta \beta_s + \Delta \beta_i,
\label{eq: JSA4}
\end{aligned}
\end{equation}
which means that all the signal (idler) Schmidt modes have the same linear phase factor defined by $\exp\left[i \frac{\Delta\beta_s L}{2}\right]$ ($\exp\left[i \frac{\Delta\beta_i L}{2}\right]$)  and, therefore, 
this common phase can be factor out in Eqs. (\ref{eq: Psi_LO}) and (\ref{eq: varphi_LO}):

\begin{equation}
u_n(\omega_s) = \tilde{u}_n(\omega_s) \exp\left[i \frac{\Delta\beta_s L}{2}\right], 
\label{eq: Signal mode}
\end{equation}
and similarly for the idler photon. Here, $\tilde{u}_n(\omega_s)$ corresponds to the decomposition of the first two terms in Eq. (\ref{eq: JSA1}). 
The obtained Schmidt modes were used to evaluate the total amount of multimode squeezing in the produced PDC radiation.

In the experiment, the theoretically calculated modes were mimicked by applying the pulse-shaping technique to the LO. The LO mode was set in one of the three Hermite-Gaussian ($HG$) modes ($HG_{0}$, $HG_{1}$ and $HG_{2}$) by implementing the set of phase masks induced by the SLM. 
The shaped spectra were measured using an optical spectrum analyzer (Thorlabs OSA 202C) and are shown in Fig. \ref{fig: Schmidt modes}(c) by the black curves. Here, the red curves represent the fits by the corresponding Hermite–Gaussian functions. Note that even after shaping the considered LO modes do not coincide with any of the Schmidt modes due to the large discrepancy between the LO spectral widths ($330\ \text{pm}$) and PDC width ($60\ \text{nm}$). 
 Even for the type-II PDC process, the width of the narrowest Schmidt mode (according to theoretical calculations) is about $1.53\ \text{nm}$, which is five times greater than the spectral width of the LO mode.

\subsection{Multimode squeezing}

\begin{figure*}[t]
  \centering
  \includegraphics[width=1.0\textwidth]{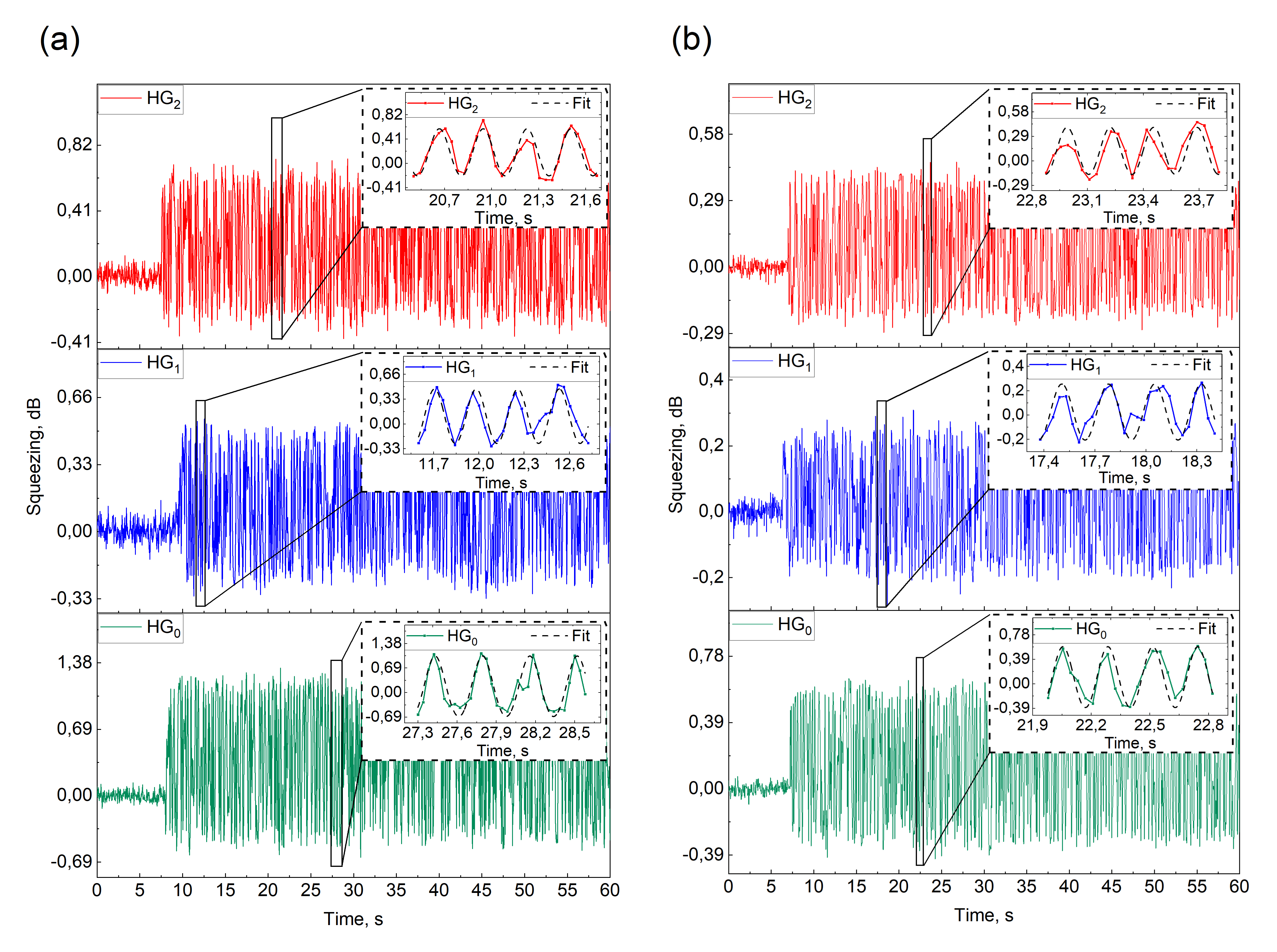}
  \caption{Measured squeezing and anti-squeezing in the first three PDC modes defined by the spectral shape of the LO: a) type-0 PDC; b) type-II PDC. The inserts show a zoom of the marked regions with a fit calculated using Eq. (\ref{squeezing}) together with Eq. (\ref{eq: Q_var1}) and Eq. (\ref{eq:Q_var type-2}) for the type-0 and type-II PDC processes, respectively. For the fit  we supposed $\phi= a* t + b$, where $a$ and $b$ are the fitting parameters, $t$ is time. The gain parameter $\Gamma$ was also considered as a fitting parameter.
  \label{fig: Squeezing}}
\end{figure*}

To measure squeezing, the power of the LO  was fixed at $P=5\ \text{mW}$ and  the periodically sweeping of the piezomirrow (PM) was performed to change the phase of the LO. The variance of the quadratures for both the type-0 and type-II PDC was measured depending on the LO phase and  normalized to the vacuum noise according to  Eq. (\ref{squeezing}). The measured squeezing is presented in Fig. \ref{fig: Squeezing} for both PDC types. The smaller values of squeezing and anti-squeezing for the type-II PDC compared to the type-0 PDC are explained by the lower efficiency of the former process. The graph also shows insets that demonstrate a zoom of the  periodic squeezing dependence on phase for both processes fitted by Eq. (\ref{squeezing}) with the use of Eq. (\ref{eq: Q_var1}) and Eq. (\ref{eq:Q_var type-2}) for the type-0 and type-II PDC, respectively. 

To obtain a fit, we performed numerical simulations using the experimentally generated LO profiles. In these simulations, the coupling strength $\Gamma$  was varied to achieve the best overlap with the experimental data in each particular case and reads $\Gamma_{HG_0}=3.61$, $\Gamma_{HG_1}=3.95$, $\Gamma_{HG_2}=3.23$ for type-0 PDC and $\Gamma_{HG_0}=0.51$, $\Gamma_{HG_1}=0.60$, $\Gamma_{HG_2}=0.61$ for type-II PDC. One can observe that the experimental results and theoretical simulations show a good agreement; some discrepancy between them is associated with the phase failures on the piezo mirror. Note that for type-II PDC, both in theory and experiment,  we used the same amplitude of the LO spectral profile for the signal and idler photons,  which is supported by calculations presented in Fig. \ref{fig: Schmidt modes} (b).

\begin{table}
\caption{\label{table1} Experimentally measured (Exp) and theoretically calculated (Theor) values of squeezing (Sq) and anti-squeezing (ASq) in dB in first three PDC modes defined by the spectral shape of LO for the type-0 and type-II PDC processes. The chosen amount of losses for theoretical calculations is shown in brackets.}
\vspace{2mm}
\centerline{
\begin{tabular}{c|c|c|c|c|c|c}
\hline
\multicolumn{7}{c}{Type-0}\\
\hline
 & \multicolumn{2}{c|}{Exp} & \multicolumn{2}{c|}{Theor (70 \%)} & \multicolumn{2}{c}{Theor (0 \%)}\\
\hline
Mode & Sq & ASq  & Sq & ASq  & Sq & ASq\\
\hline
$HG_{0}$ & -0.67 & 1.03 & -0.64 & 1.09 & -2.64 & 2.91\\
$HG_{1}$ & -0.26 & 0.47 & -0.26 & 0.41 & -0.93 & 1.24\\
$HG_{2}$ & -0.21 & 0.59 & -0.26 & 0.53 & -0.93 & 1.55\\
\hline
\multicolumn{7}{c}{Type-II}\\
\hline
& \multicolumn{2}{c|}{Exp} & \multicolumn{2}{c|}{Theor (63 \%)} & \multicolumn{2}{c}{Theor (0 \%)}\\
\hline
Mode & Sq& ASq & Sq& ASq & Sq& ASq\\
\hline
$HG_{0}$ & -0.38 & 0.59 & -0.47 & 0.59 & -1.39 & 1.45\\
$HG_{1}$ & -0.20 & 0.26  & -0.20 & 0.23  & -0.55 & 0.60\\
$HG_{2}$ & -0.16 &  0.40 & -0.22&  0.29 & -0.62&  0.73\\
\hline
\end{tabular}}
\end{table}

From  Fig. \ref{fig: Squeezing} we estimate the average experimentally measured values of squeezing and anti-squeezing for the first three HG modes of the LO, the results are presented in Tab. \ref{table1}.
In addition, Tab. \ref{table1} shows theoretically calculated squeezing (anti-squeezing) values without losses and taking into account the estimated amount of losses (detection and optical) for both processes. We consider the detection efficiency to be  $70\%$ as reported in \cite{fedorov2017nonlocalcontrolquantumstate} for a similar homodyne detector. Optical losses were estimated as $13\%$. An additional amount of losses is present due to the adjustment of the polarization between the LO and the PDC during the detection process and is estimated as $50\%$ for type-0 and $40\%$ for type-II PDC, resulting in the total efficiency of  the entire system of 0.30 (or $70\%$ losses) for type-0, and 0.37 (or $63\%$ loses)  for type-II PDC.

For the theoretical calculation presented in Tab. \ref{table1}, the coupling strength $\Gamma$ was set to $\Gamma=3.67$ for the type-0 and $\Gamma=0.56$ for the type-II PDC to achieve the best overlap with the experimental data.
Taking into account the first eigenvalues for both types - $\lambda_0=0.0135$  and $\lambda_0=0.1402$, respectively - the experimental gain  can be estimated as $G=\sqrt{\lambda_0} \Gamma $ and equals to G=0.43 (G=0.21) for type-0 (II) PDC.

\subsection{Reconfigurable cluster states}

\begin{figure}[!t]
\centering
\includegraphics[width=\columnwidth]{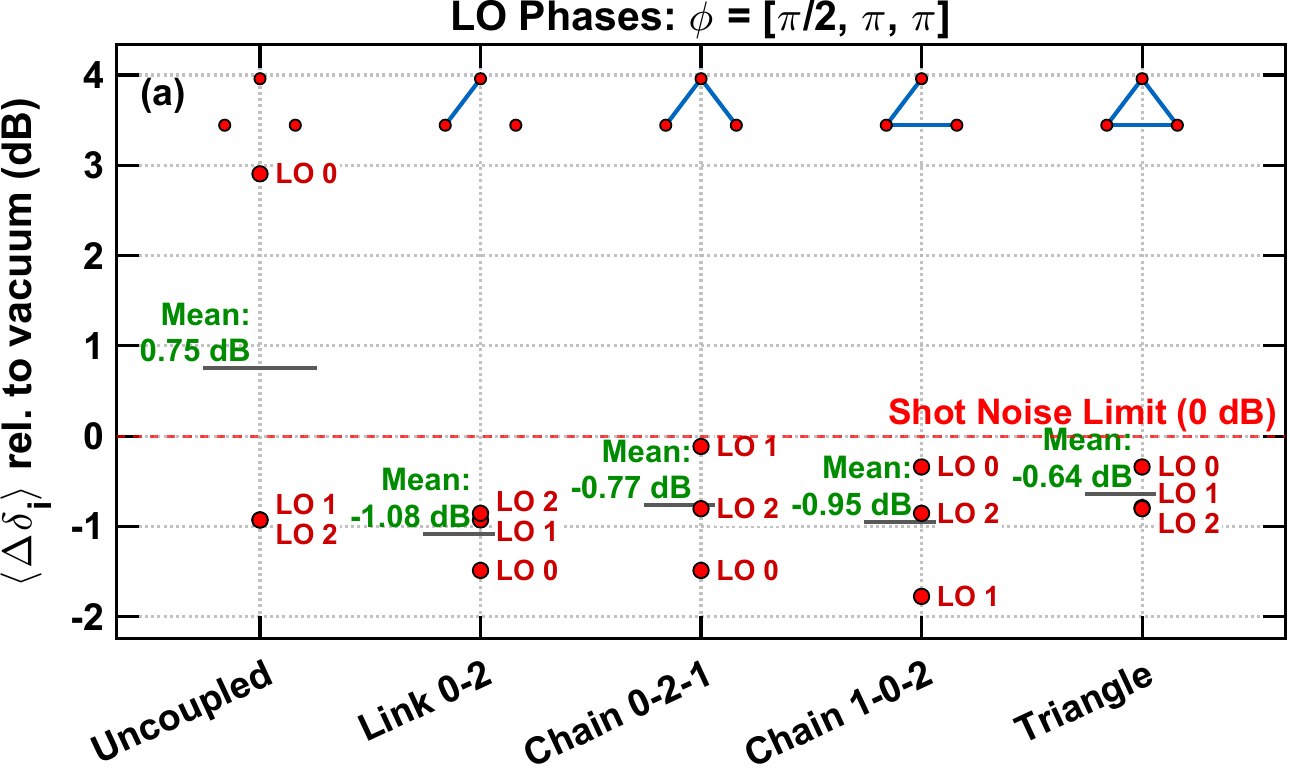}\\[+1mm]
\includegraphics[width=\columnwidth]{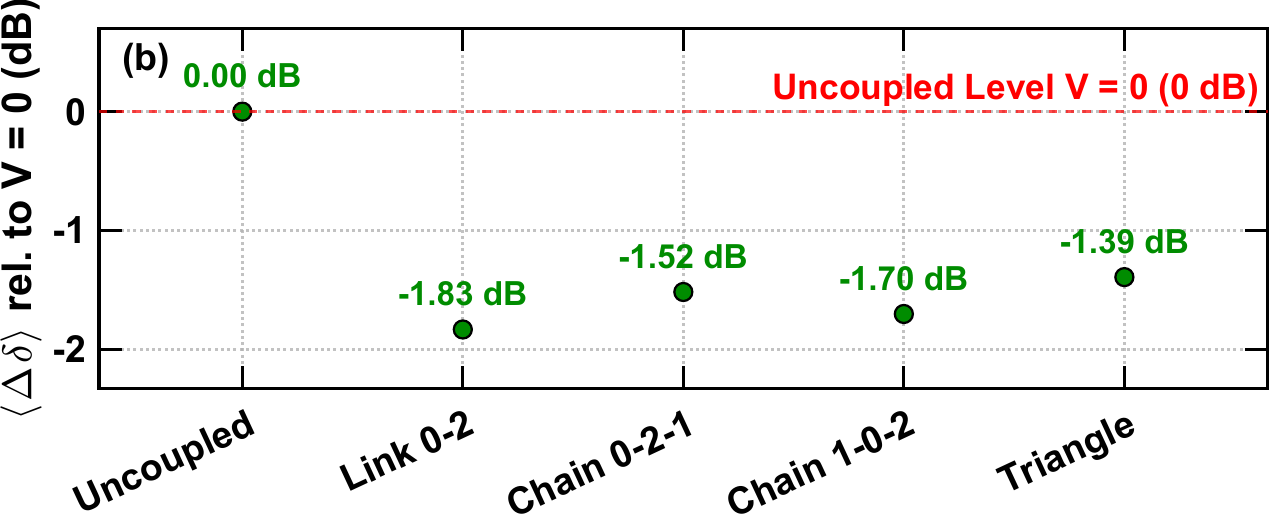}
\caption{
 Squeezing levels (in decibels) of cluster nullifiers $\langle \Delta \hat{\delta}_i \rangle$ for $\phi = [\pi/2, \, \pi, \, \pi]$. 
    (a) Squeezing relative to the vacuum shot-noise level  defined for each nullifier as $\langle \Delta \hat{ \delta}_i \rangle_{\text{vac}} = \frac{1}{4}(1 + \sum_j V_{ij}^2)$.
Red circles denote individual LO modes; horizontal gray lines indicate the arithmetic mean of the linear noise  converted to decibels.
(b) Node-averaged squeezing (green circules) relative to the uncoupled state ($V_{ij} = 0$).
}
\label{fig:cluster_results}
\end{figure}

One of the key applications of the considered multimode squeezed light is the deterministic generation of continuous-variable cluster states, which serve as a universal resource for measurement-based quantum computing \cite{vanLoock2007, PhysRevA.68.022312, Hasnaoui_2026, feng2026onchipgenerationmultiqubitgraph}. In the case of multimode squeezed light, by shaping the LO spectral profile, various reconfigurable cluster states can be certified without complex linear optical networks \cite{PhysRevResearch.6.043113}. The architecture of a cluster state can be described by a graph with a uniquely defined adjacency matrix $V_{ij}$, setting the nullifier operators $\hat{\delta}_i = \hat{P}_i - \sum_j V_{ij}\hat{Q}_j$ \cite{Menicucci2011}, where $\hat{Q}_j= \frac{ \hat{A}_{LO}^{(j)} + (\hat{A}_{LO}^{(j)})^{\dagger}}{2}$ and $\hat{P}_i=\frac{ \hat{A}_{LO}^{(i)} -(\hat{A}_{LO}^{(i)})^{\dagger}}{2i}$ are the coordinate and momentum quadratures of multimode squeezed light projected by the LO modes. Below we consider four representative graph topologies (shown schematically in Fig.~\ref{fig:cluster_results}): the uncoupled state (all $V_{ij} = 0$), the bipartite link $V_{0-2}$ (with $V_{02}=1$ and all other $V_{ij} = 0$), the tripartite linear chains $V_{0-2-1}$ (with $V_{01}=0$ and all other $V_{ij} = 1$) and $V_{1-0-2}$ (with $V_{12}=0$ and all other $V_{ij} = 0$), and the fully connected triangle (with $V_{ij} = 1 - \delta_{ij}$). 
In type-0 PDC, expanding the LO modes over the Schmidt basis via the coefficients $C_n^k = M_n^k e^{i\theta_n^k} = \alpha_n^k + i\beta_n^k$, where the superscript $k \in \{0, 1, 2\}$ denotes the LO mode shaped into the $HG_k$ mode (generalizing Eq.~(\ref{eq: Psi_LO})), while $\alpha_n^k$ and $\beta_n^k$ represent the real and imaginary parts,  the exact variance of the $i$-th nullifier reads:
\begin{equation}
\begin{aligned}
\langle \Delta \hat{\delta}_i \rangle = & \sum_{n=0}^\infty \left[ \frac{e^{-2r_n}}{4} \left( \alpha_n^i - \sum_j V_{ij}\beta_n^j \right)^2 \right.
\\[2mm]
& + \left. \frac{e^{2r_n}}{4} \left( \beta_n^i + \sum_j V_{ij}\alpha_n^j \right)^2 \right],
\label{eq:nullifier_var}
\end{aligned}
\end{equation}
where we take into account that $\langle \hat{\delta}_i \rangle =0 $ for the squeezed vacuum light.

To analyze the effect of the LO phase control, we introduce programmable phase shifts $\phi_k$, setting the rotated quadratures to $\tilde{\alpha}_n^k = M_n^k \cos(\theta_n^k + \phi_k)$ and $\tilde{\beta}_n^k = M_n^k \sin(\theta_n^k + \phi_k)$. For any coupled pair of nodes $(i, j)$ with $V_{ij} = 1$, the sum of their anti-squeezed contributions ($\propto e^{2r_n}$) for each Schmidt mode $n$ in Eq.~(\ref{eq:nullifier_var}) expands as:
\begin{equation}
\label{eq:anti_squeezed_sum}
\begin{aligned}
(\tilde{\beta}_n^i + \tilde{\alpha}_n^j)^2 + (\tilde{\beta}_n^j + \tilde{\alpha}_n^i)^2 = 
(M_n^i)^2 + (M_n^j)^2 \\
+ 2M_n^i M_n^j \sin\left(\phi_i + \phi_j + \theta_n^i + \theta_n^j\right).
\end{aligned}
\end{equation}
In particular, for the bipartite link between $HG_0$ and $HG_2$ ($i=0, j=2$),
the signs match, namely,
 $\theta_n^0 = \theta_n^2 \in \{0, \pi\}$, so their  sum satisfies $\theta_n^0 + \theta_n^2 \equiv 0 \pmod{2\pi}$. Minimizing Eq.~(\ref{eq:anti_squeezed_sum}) over LO phases then yields the condition $\sin(\phi_0 + \phi_2) = -1$ that results in $\phi_0 + \phi_2 = 3\pi/2 \pmod{2\pi}$. In this case, the anti-squeezed noise reduces to $(M_n^0 - M_n^2)^2$. A systematic optimization of Eq.~(\ref{eq:anti_squeezed_sum})  over the LO phases reveals that the configuration $\phi = [\pi/2, \, \pi, \, \pi]$ provides optimal noise suppression across all considered topologies.  
This compensation relies on the spectral mismatch between the LO and PDC modes, which ensures that different LO modes share common Schmidt modes ($M_n^i M_n^j \neq 0$). If the LO modes matched isolated Schmidt modes, this cross-term would vanish ($M_n^i M_n^j = 0$), leaving the anti-squeezed noise uncompensated.

To assess the capability of the source, we evaluate Eq.~(\ref{eq:nullifier_var})  in the absence of losses  ($R = 0$) using theoretically calculated eigenmodes and eigenvalues, which corresponds to the squeezing values presented in the Theor (0\%) column in Tab. \ref{table1}. As shown in Fig.~\ref{fig:cluster_results}(a), the selected phase setting yields squeezing below the vacuum limit across the all tested topologies. One can notice that in the uncoupled case ($V_{ij} = 0$), the anti-squeezed $HG_0$ mode ($+2.91$~dB) results in an elevated average baseline $+0.75$~dB. Upon engaging graph connectivity, this excess noise is coherently reduced, leading to the net enhancement. For the bipartite link $V_{0-2}$, the average nullifier squeezing reaches  $-1.08$~dB below the vacuum limit (Fig.~\ref{fig:cluster_results}(a)) and an enhancement of $-1.83$~dB over the uncoupled state (Fig.~\ref{fig:cluster_results}(b)). For the tripartite linear chains $V_{0-2-1}$ and $V_{1-0-2}$, the average nullifier squeezing reaches $-0.77$~dB and $-0.95$~dB ($-1.52$~dB and $-1.70$~dB over the uncoupled state), respectively. Even for the fully connected triangle, the squeezing reaches $-0.64$~dB ($-1.39$~dB over the uncoupled state). 
These results show that varying the spectral shape of the LO allows a single-pass type-0 PDC to serve as a viable platform for generating continuous-variable cluster states by tailoring the LO profiles and phases.

\section{Conclusion}

In this work, we characterized both theoretically and experimentally the source of multimode squeezed light based on type-0 and type-II PDC processes in a PPKTP crystal under picosecond pumping. We performed numerical simulations based on the Schmidt-mode theory to estimate the amount of measured squeezing depending on the phase and spectral profiles of the local oscillator.
We then experimentally proved our assumptions by measuring the squeezing of the first three Hermite-Gaussian modes using the phase- and spectrally-shaped local oscillator. The results show good agreement between the theory and the experiment. In addition, we considered different cluster state topologies corresponding to the given experimental conditions  and analyzed their properties.  Finally, we analyzed the impact of losses on the mode squeezing.  
The presented approach, which takes into account the parameters of both the source and the local oscillator, could play an important role in future experiments with multimode squeezed light.

\section{Acknowledgments}
We acknowledge the financial support of the RSF grant No. 26-12-00198, E.M. personally thanks the  ‘BASIS’ Foundation support (project №26-2-2-45-1).

\appendix

\section{Losses}
\label{app: C}

Linear losses can be modeled by a fictional beam splitter with reflection coefficient  $R$ that mixes the mode of interest  $\hat a$ with a vacuum mode $\hat{a}_{vac}$. Then, the output light reads

\begin{equation}
\hat a_{\text{loss}} = \sqrt{1-R} \cdot \hat a_{\text{}} + \sqrt{R} \cdot \hat{a}_{vac}.
\label{eq: a_loss}
\end{equation}

For the quadrature operator  $\hat Q = \frac{1}{2}\left(\hat a e^{-i\theta} + \hat a^\dagger e^{i\theta}\right)$, the corresponding output quadrature accounting losses is

\begin{equation}
\hat Q_{\text{loss}} =\sqrt{1-R} \cdot \hat Q  +  \sqrt{R} \cdot \hat Q_{vac}.
\label{eq: Q_loss}
\end{equation}

Since the vacuum mode is uncorrelated with the initial field, $[ \hat Q, \hat Q_{vac}] = 0,$ the output variance is given by

\begin{equation}
\langle \hat Q_{\text{loss}}^2\rangle = (1-R)\cdot \langle \hat Q^2 \rangle +  R \cdot \langle \hat Q_{\text{vac}}^2 \rangle.
\label{eq: Q^2_loss}
\end{equation}

\section{Pulse shaping}
\label{app: D}
\begin{figure*}%
  \centering \includegraphics[width=1\textwidth]{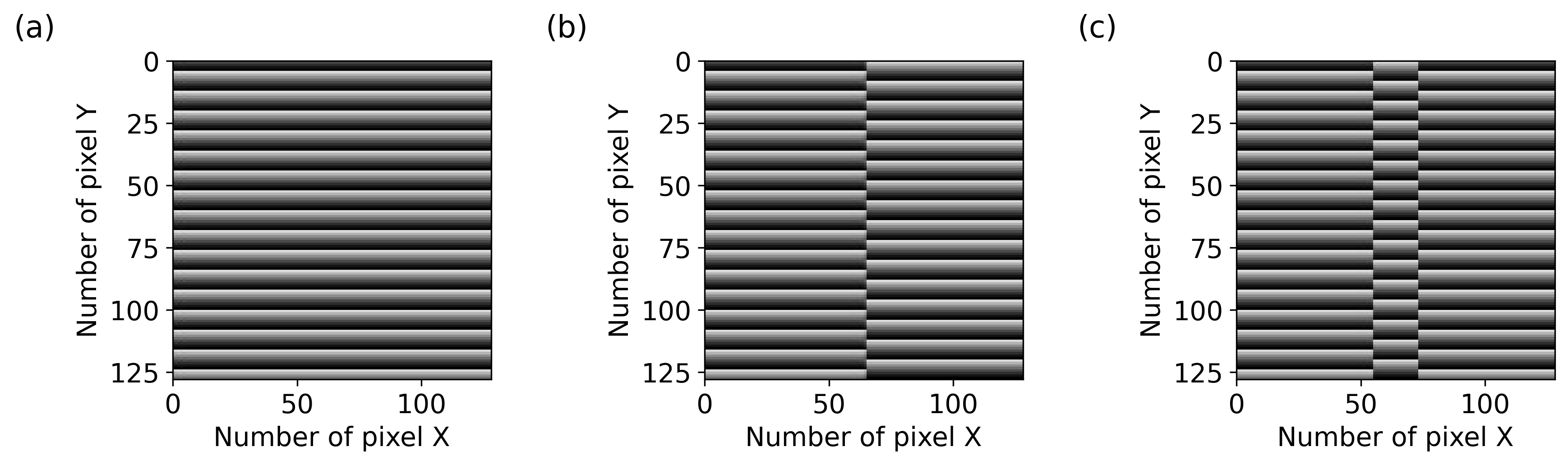}
  \caption{Phase masks for the first three Hermite-Gauss modes: a) $HG_{0}$; b) $HG_{1}$; c) $HG_{2}$.
  \label{fig: Phase masks}}
\end{figure*} 

\begin{figure*}[t]
  \centering  \includegraphics[width=1.0\textwidth]{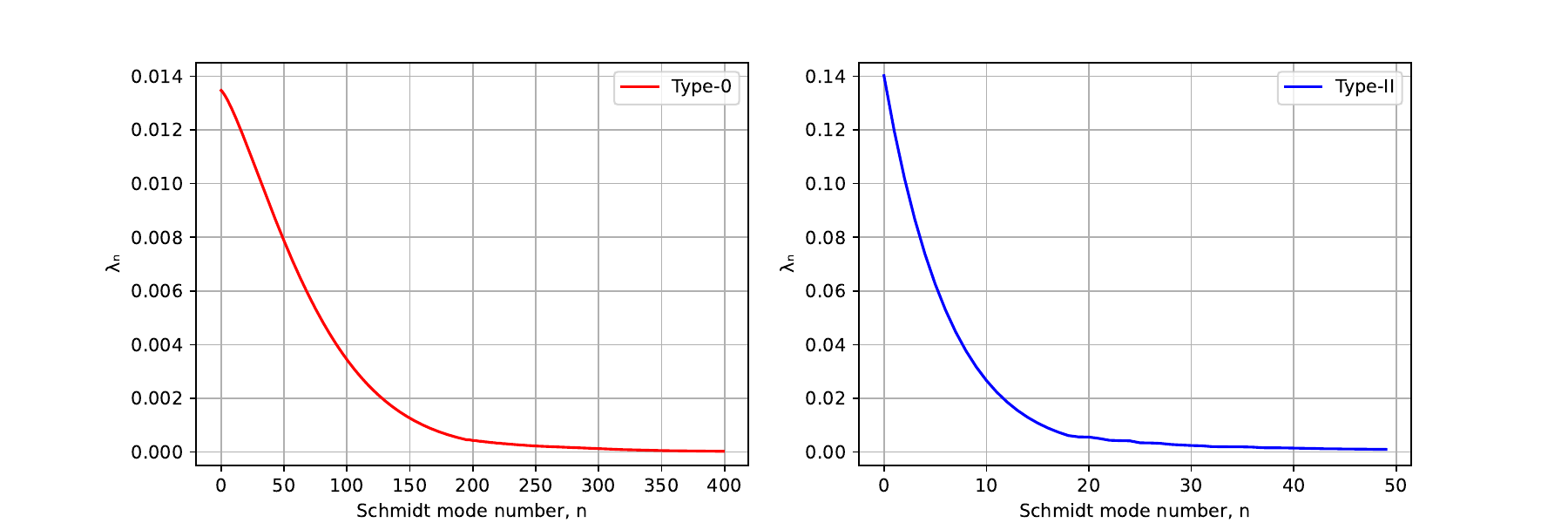}
  \caption{Theoretical eigenvalue distributions in the case of (a) type-0 PDC and (b) type-II PDC processes. }.
  \label{fig: eigvals}
\end{figure*}

Phase masks were generated for the first three Hermite-Gauss modes following the procedure described in \cite{Kouadou23}, where, depending on the displacement of the horizontal alternating gray level bands, a certain phase delay was applied to certain blocks of the SLM matrix. Examples of generated phase masks are shown in Fig. \ref{fig: Phase masks}.

\section{Eigenvalue distributions} \label{appendix: Schmidt eigenvalues}

The Schmidt eigenvalues $\lambda_n$ were calculated using Eq. \eqref{eq: JSA schmidt decomposition} and are presented in Fig. \ref{fig: eigvals} for both types. It is clearly seen that for the type-0 PDC process,
the number of Schmidt modes is significantly larger (by almost an order of magnitude) than for its type-II analogous.

\bibliography{Ref}

\end{document}